\documentclass[twocolumn,showpacs,prl,amsmath,amssymb,floatfix]{revtex4}
\usepackage{amsmath,amssymb}
\usepackage{bm}
\usepackage{epsfig}
\usepackage{psfrag}

\newcommand{\bd}{\bm}

\begin{document}

\title{Absence of fermionic quasi-particles in the superfluid state of the 
attractive Fermi gas}

\author{Nils Lerch, Lorenz Bartosch, and Peter Kopietz}
  
\affiliation{Institut f\"{u}r Theoretische Physik, Universit\"{a}t
  Frankfurt,  Max-von-Laue Strasse 1, 60438 Frankfurt, Germany}

\date{October 17, 2007}

 \begin{abstract}
We calculate  the effect of order parameter fluctuations 
on the fermionic single-particle excitations in the superfluid state
of   neutral fermions  interacting with  short range attractive forces.
We show that in dimensions $D \leq 3$ 
the singular effective interaction between the fermions mediated by the
gapless Bogoliubov-Anderson mode prohibits the existence of
well-defined quasi-particles.
We explicitly calculate the single-particle spectral function
in  the BEC regime in $D=3$ and show that 
in this case the quasi-particle residue and the density of states
are logarithmically suppressed. 
\end{abstract}


\maketitle

Recent  experimental progress in cooling atomic Fermi gases to
ultracold temperatures \cite{Bloch07} has revived the interest 
in the BCS-BEC crossover in the attractive two-component Fermi gas.
Although a simple mean-field approximation
is sufficient to obtain a qualitatively correct description of  the 
thermodynamics of this crossover \cite{Eagles69,Leggett80}, 
for comparison with experiments  numerically accurate calculations beyond
the mean-field approximation are necessary. 
Fluctuation corrections to the free energy have been calculated
at the level of the ladder approximation  \cite{Nozieres85}, which 
 in the functional  integral approach is equivalent 
to the Gaussian approximation in the order parameter fluctuations \cite{Drechsler92,Melo93,Randeria95,Engelbrecht97,Alexandrov93,Stoof93,Palo99,Hu06,Veillette06,Nikolic07,Diener07}.
Close to the unitary point, where the $s$-wave scattering length is large, 
the Gaussian approximation is not 
sufficient and more sophisticated many-body methods are 
necessary \cite{Haussmann07,Diehl07}.

While most authors focused on the {\it{thermodynamics}} of the BCS-BEC crossover,
not much attention has been payed to the {\it{fermionic single-particle excitations}}
in the BEC regime.
Fluctuation corrections to the
fermionic self-energy have been briefly considered  in the appendix of
Ref.~[\onlinecite{Veillette06}],  but   the momentum and 
frequency dependent  single-particle spectral function $\rho ( \bd{k} , \omega )$
and the density of states $\nu ( \omega )$ have not been analyzed. 
In this Letter we shall show that  in dimensions $ D \leq 3$
the singular interaction between {\it{neutral}} fermions mediated by the gapless
Bogoliubov-Anderson (BA) mode  
leads to the breakdown of the quasi-particle picture.
We emphasize that our result 
is relevant for superfluid  gases of cold fermionic atoms, but 
does not apply to the superconducting state of charged fermions, where
the long-range Coulomb interaction pushes the BA mode up
to high energies~\cite{Schrieffer83}.

For our explicit  calculations of 
$\rho ( \bd{k} , \omega )$
and  $\nu ( \omega )$ we shall focus on the BEC regime in
the physically most relevant dimension $D=3$ at zero temperature.
It turns out that in this case
the imaginary part of the self-energy vanishes
linearly in the properly shifted  frequency variable $y$ (see below), 
implying that the frequency dependent real part
of the self-energy vanishes as $y \ln y$.
A similar frequency dependence of the self-energy
in the normal state of strongly correlated electrons
has been proposed phenomenologically  by
Varma {\it{et al.}}~\cite{Varma89} to explain the unusual normal state properties
of the high-temperature superconductors.
While a generally accepted theory explaining
the microscopic  origin of such a marginal Fermi liquid behavior 
in the {\it{normal}} state
does not exist, we shall give here  a microscopic derivation of similar
marginal quasi-particle behavior
of  the fermionic single-particle excitations  in the
{\it{superfluid}} state of neutral fermions in 
three dimensions.

To begin with, let us briefly outline the derivation of
fluctuation corrections to the fermionic self-energy {\mbox{using}} functional integration.
We consider a system of neutral fermions with dispersion 
$\epsilon_{\bd{k}} = \bd{k}^2  / (2m)$ and a short-range  attractive
two-body interaction $g_{\bd{p}} > 0$ depending on the 
total two-particle momentum $\bd{p}$.
Writing the fermionic single-particle Green function
$\bf{G} $ in the Nambu-Gorkov basis~\cite{Schrieffer83} 
as a  functional integral and decoupling
the interaction in the spin-singlet particle-particle channel
by means of a complex bosonic Hubbard-Stratonovich (HS) field $\psi$,
we can represent $\bf{G} $ as a functional average,
\begin{eqnarray}
 {\bf{G}} 
 & = &
 \frac{ \int {\cal{D}}[ \delta \bar{\psi} , \delta \psi] 
      e^{-  S_{\rm eff} [ \delta \bar{\psi} , \delta \psi ] }   {\mathbf{G}}_{\psi}     }{
\int {\cal{D}}[  \delta \bar{\psi} , \delta \psi] 
 e^{- S_{\rm eff} [ \delta \bar{\psi} , \delta \psi ] } },
 \label{eq:Gexactdef} 
\end{eqnarray}
where ${\bf{G}}_{\psi} $ is the fermionic Green function for a fixed configuration
of the HS field.  Its inverse is given by the following
matrix in Nambu-Gorkov and  momentum-frequency space \cite{footnotelabels},
 \begin{equation}
 [ {\bf{G}}_{\psi}^{-1} ]_{ K K^{\prime}} = 
  \left[ \begin{array}{cc}
 \delta_{ K , K^{\prime} } ( i \omega - \xi_{ \bd{k}} ) &
 - \psi_{ K - K^{\prime} }\\
- \bar{\psi}_{ K^{\prime} - K} &   \delta_{ K , K^{\prime} } ( i
\omega + \xi_{ - \bd{k} } ) 
 \end{array}
 \right] .
 \label{eq:Gdef}
 \end{equation}
Here  $\xi_{\bd{k}} =
\epsilon_{\bd{k}} - \mu$ where $\mu$ is  the chemical potential, and
the effective action $S_{\rm eff} [ \delta \bar{\psi} , \delta \psi ]$ of the HS field is
 \begin{equation}
 S_{\rm eff} [ \delta \bar{\psi} , \delta \psi ]  =  
 \int_P g_{\bd{p}}^{-1} \delta \bar{\psi}_P \delta \psi_P
 + \sum_{n=2}^{\infty} \frac{
 {\rm Tr} [ {\bf{G}}_{0}  \mathbf{V}  ]^n}{n},
 \label{eq:Sefffinal}
 \end{equation}
where $\delta \psi_P = \psi_ P -  \delta_{ P,0} \Delta_0$, and the
mean-field Green function ${\bf{G}}_{0}$  
is obtained from Eq.~(\ref{eq:Gdef})
by approximating  the HS field by its saddle point $\psi_P \approx \delta_{ P,0} \Delta_0$. 
The fluctuation matrix $ \mathbf{V}$ is defined by
$    \mathbf{V} =   {\mathbf{G}}_{0}^{-1} -  {\mathbf{G}}_{\psi}^{-1}  $.
The saddle point condition
leads  to the usual BCS gap equation for  $\Delta_0$ and the quasi-particle
dispersion $E_{\bd{k}} = \sqrt{ \xi_{\bd{k}}^2 + \Delta_0^2}$.
For convenience we neglect the momentum dependence of
$g_{\bd{p}}  \approx  g_0$ and regularize the resulting
ultraviolet divergence \cite{Randeria95}   in the BCS gap equation 
by eliminating $g_0$ in favor of
the corresponding two-body $T$-matrix in vacuum, which we denote by  $g$.
The relevant dimensionless interaction is then
$ \tilde{g} = \nu_0 g $, where $\nu_0$ is the
density of states at the Fermi energy in the absence of interactions.
In $D=3$ we have
 $\tilde{g} = -  2 k_F  a_s / \pi$, where $a_s$ is the
$s$-wave scattering length in vacuum and $k_F = mv_F = \sqrt{ 2 m \epsilon_F}$ 
is the Fermi momentum.

\begin{figure}[tb]
  \centering
  \epsfig{file=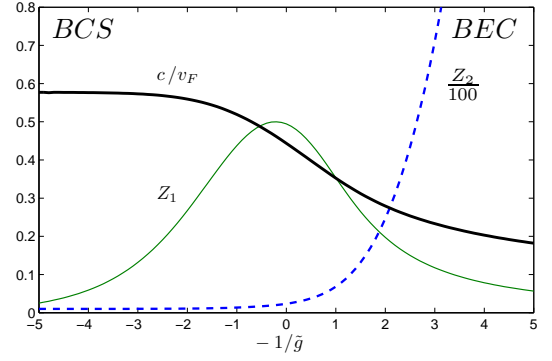,width=70mm}
  \vspace{-4mm}
  \caption{%
(Color online)
Factors $Z_1$,
${Z}_2$, and $c/v_F$ defined in the text
as a function of
$- 1/ \tilde{g} = \pi /( 2 k_F  a_s)$ for $D=3$.
}
  \label{fig:Zfactors}
\end{figure}

The exact fermionic propagator
$\mathbf{G}$ defined in Eq.~(\ref{eq:Gexactdef}) can be written as
$ {\mathbf{G}} = [ {\bf{G}}_{0}^{-1} - \mathbf{\Sigma} ]^{-1}$,
where the self-energy matrix  $\mathbf{\Sigma} $  
involves a normal component $\Sigma (K)$ and
an anomalous component $\delta \Delta ( K )$ (which can be viewed as a fluctuation
correction to $\Delta_0$),
\begin{eqnarray}
 [ \mathbf{\Sigma}]_{ K K^{\prime}} = \delta_{ K , K^{\prime}} 
  \left[ \begin{array}{cc}
  \Sigma ( K ) &
  \delta  \Delta ( K ) \\
 \delta \Delta^{\ast} (K ) &   
 -  \Sigma (-K ) 
 \end{array}
 \right].
 \label{eq:Sigmamatrix}
 \end{eqnarray}
The leading contribution to  $\mathbf{\Sigma}$ can be written as
  $\mathbf{\Sigma} = \langle \mathbf{V} \mathbf{G}_0 \mathbf{V}  \rangle$,
where $\langle \ldots \rangle$ denotes 
functional average with the 
effective action  (\ref{eq:Sefffinal}) in
Gaussian approximation, retaining only quadratic terms in the fluctuations.
As emphasized by Castellani {\it{et al.}}~\cite{Castellani97} (see also
Ref.~\cite{Kreisel07}),
the scaling behavior of the
order parameter correlation functions is
more transparent if
we  express the complex field $\delta \psi_P$
in terms of two real fields \cite{footnotereal} by setting
 $\delta \psi_P =  [ \chi_P + i \phi_P]/ \sqrt{2}$ and
 $\delta \bar{\psi}_P =  [ \chi_{-P} - i \phi_{-P}]/ \sqrt{2}$.
The fields $\chi_P$ and $\phi_P$ correspond to
longitudinal and transverse fluctuations of the superconducting 
order parameter, respectively.
The effective action (\ref{eq:Sefffinal}) in Gaussian approximation
is then  \cite{footnotelabels}
\begin{equation}
 S_{\rm eff} [ \chi , \phi ] \approx  \frac{1}{2} \int_P 
( \chi_{P}^{\ast} , \phi_{P}^{\ast} ) 
 \left[ \begin{array}{cc}
 F^{\chi \chi}_P  & F^{ \chi \phi}_P  \\
F^{\phi  \chi }_P   & F^{\phi \phi}_P 
 \end{array} \right]^{-1}
\left( \begin{array}{c} \chi_{P} \\ \phi_{P} \end{array}
 \right),
 \end{equation} 
where 
for small $ \bd{p}$ and $\bar{\omega}$ 
the matrix elements of the bosonic propagator are
 \begin{subequations}
\begin{eqnarray}
 F_P^{\chi \chi } & \approx &  \frac{ D c^2 }{ \nu_0 v_F^2} \frac{ \bar{\omega}^2 
+ Z_2 c^2 \bd{p}^2  }{
 \bar{\omega}^2 + c^2 \bd{p}^2 },
 \label{eq:Fchichilong}
 \\
 F_P^{\phi \phi } & \approx &  \frac{ D c^2}{ \nu_0 v_F^2} \frac{ ( 2 \Delta_0 )^2}{
 \bar{\omega}^2 + c^2 \bd{p}^2 },
  \label{eq:Fphiphilong}
 \\
 F_{P}^{\chi \phi } = - F_{P}^{\phi \chi }  & \approx &    \frac{ Z_1   }{ \nu_0} 
 \frac{  2 \Delta_0    \bar{\omega} }{
 \bar{\omega}^2 + c^2 \bd{p}^2 }.
  \label{eq:Fchiphilong}
 \end{eqnarray}
 \end{subequations}
Introducing the dimensionless integrals 
 \begin{eqnarray}
I_1 & = & 
\frac{1}{ \nu_0 V  } \sum_{\bd{k}} \frac{ \Delta_0^2}{ 2 E_{\bd{k}}^3}   , 
 \; \; \; 
 I_2  =
\frac{1}{ \nu_0 V } \sum_{\bd{k}} \frac{   \Delta_0 \xi_{\bd{k}} }{  2 E_{\bd{k}}^3},
 \label{eq:Z1def}
\end{eqnarray}
we have
$Z_1 = I_1 I_2 / [ I_1^2 + I_2^2]$ and $Z_2 
= [ I_1^2 + I_2^2]/ I_1^2 $, while 
the velocity $c$ of the BA mode can be written 
as $D c^2 / v_F^2 = I_1 / [ I_1^2 + I_2^2]$.
Numerical results for $Z_1$, $Z_2$ and $c/v_F$ 
in $D=3$ as a function of the relevant dimensionless 
coupling $- \tilde{g}^{-1}$
are shown in Fig.~\ref{fig:Zfactors}.
In $D$ dimensions we obtain  in the BCS limit $c \approx v_F / \sqrt{D}$,
and in the BEC limit $c = \sqrt{ \rho_{B} U_{\ast} / m_B}$, with the
effective interaction $ U_{\ast} = \frac{4-D}{D-2} |g|$.
Here
$\rho_B = \rho/2$ and $m_B=2m$
are the density and the mass of bosons formed by  paired
fermions with density $\rho$ and mass $m$.
Our result for $c$ in the BEC limit
agrees with the well-known Hartree-Fock result for the velocity
of the elementary excitations in the interacting Bose gas provided
we identify $U_{\ast}$ with the effective Hartree-Fock 
potential.
Note that only in three dimensions $U_{\ast}$ is given by
the zero-energy $T$-matrix $|g|$ of the underlying fermionic system.
In the limit $D \rightarrow 2$ the $T$-matrix vanishes as
 $  g   \sim - (D-2) / \nu_0$
(see Ref.~[\onlinecite{Sauli06}]), so that in two dimensions
$U_{\ast} = 2 / \nu_0$ and 
hence $c  = v_F / \sqrt{2}$ for all values of $g_0$~\cite{Marini98}.
Carrying out the Gaussian average in 
 $\mathbf{\Sigma} = \langle \mathbf{V} \mathbf{G}_0 \mathbf{V}  \rangle$, we 
finally obtain for the fermionic self-energies defined in Eq.~(\ref{eq:Sigmamatrix}),
\begin{eqnarray}
 \Sigma ( K ) & = & - \frac{1}{2} \int_P [ F^{\phi \phi}_P + F^{\chi \chi}_P
 +  2 i F_P^{\chi \phi} ] B_0 ( P-K), \hspace{10mm}
 \label{eq:SigmaKGauss}
 \\
   \delta \Delta ( K ) & = & - \frac{1}{2} \int_P [ F^{\phi \phi}_P - F^{\chi \chi}_P
 ] A_0 (P- K),
 \label{eq:DeltaKGauss}
\end{eqnarray}
where
$B_0 ( K )  =  -  [ i \omega + \xi_{\bd{k}} ]/ [\omega^2 + E_{\bd{k}}^2 ]$
and $A_0 ( K )  =  -\Delta_0/ [ \omega^2 + E_{\bd{k}}^2 ]$ are the normal and
anomalous components of the mean-field propagator $\mathbf{G}_{0}$.

\begin{figure}[tb]
  \centering
  \epsfig{file=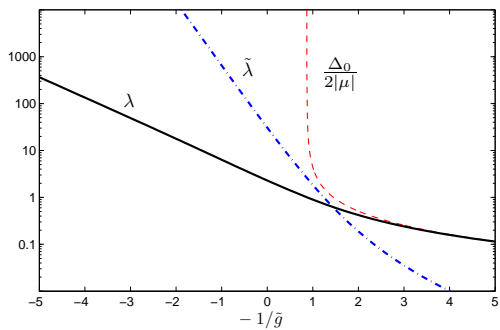,width=75mm}
  \vspace{-4mm}
  \caption{%
(Color online)
Dimensionless parameter $ \lambda= 4 mc^2 / \Delta_0$ 
and $\tilde{\lambda} = 8 \lambda^3/\pi$ in $D=3$
as a function 
of $-1 / \tilde{g}$. 
For comparison we also show the ratio $ \Delta_0 / ( 2 | \mu |)$,
which approaches $4 mc^2/  \Delta_0$ in the BEC limit.
}
  \label{fig:deltacc}
\end{figure}

It turns out that in the BEC limit, where the 
dimensionless parameter $\lambda = 4 mc^2 / \Delta_0 \approx
\Delta_0 /(2 | \mu | )$ is small compared with unity (see Fig.~\ref{fig:deltacc}),
the integrations in Eqs.~(\ref{eq:SigmaKGauss},\ref{eq:DeltaKGauss})
can be explicitly carried out in $D=3$ at zero temperature,
provided the effective interactions mediated by
the bosonic fields are approximated by Eqs.~(\ref{eq:Fchichilong}--\ref{eq:Fchiphilong}).
In this regime the
$s$-wave bound state energy
$\epsilon_s \approx 2 | \mu |$ is the largest  energy scale in the problem.
Then  we may expand the mean-field dispersion  as
 $E_{\bd{k}} \approx E_0 + \bd{k}^2 /(2 m_{\ast})$, 
with $E_0 \approx | \mu | + \Delta_0^2/ \epsilon_s = | \mu | ( 1 + 2 \lambda^2)$ and
$m / m_{\ast} \approx 1 - 2 \lambda^2$ for small $\lambda$.
In fact, the leading non-analytic behavior of the self-energies
is entirely due to the infrared singularity
in the propagator $F^{\phi \phi }_P$ 
of the BA mode in Eq.~(\ref{eq:Fphiphilong}), so that
we may neglect the contributions from $F^{\chi \phi}_P$ and
$F^{\chi \chi }_P$ in Eqs.~(\ref{eq:SigmaKGauss},\ref{eq:DeltaKGauss}).
Within this approximation, which is accurate 
for  $  | \bd{k} | a_s \ll 1$ and 
$\bigl| | \omega | - E_{\bd{k}} \bigr| \ll | \mu |  $,
we obtain
 \begin{eqnarray}
  \Sigma ( \bd{k} ,  i \omega ) & = &  - 2 \Delta_0 
 \left[ \lambda^2 C ( x , z )   -  ( 1 + \lambda^2 )  C ( x , z^{\ast} )     \right],
\hspace{7mm}
 \label{eq:selfnres}
 \\
 \delta \Delta ( \bd{k} ,  i \omega ) & = & 2 \Delta_0 
 \lambda \left[  C ( x , z ) +  C ( x , z^{\ast} ) \right]
,
 \label{eq:selfares}
 \end{eqnarray}
where  
 $ x =  | \bd{k} | /k_c$ and  $  z =  ( i \omega - E_{\bd{k}})/ \omega_c$,
with  $k_c = mc $ and
$\omega_c = mc^2/2$.
The complex function $C ( x, z )$ is 
 \begin{eqnarray}
 \pi \, C ( x , z ) & = & 2 + \ln \Bigr( \frac{ M^2 }{-   z } \Bigl)
 - \frac{1}{2x} \Bigl\{
 \sqrt{ (1+x)^2 + z }
 \nonumber 
 \\
 & & \hspace{-17mm} \times
\ln \Bigl[ \frac{ 1+ x + \sqrt{ (1+x )^2 + z }}{ 1+x  
 - \sqrt{ (1+x)^2 +z } } \Bigr] - (x \rightarrow -x ) \Bigr\},
\end{eqnarray}
where the ultraviolet cutoff $M \approx  ( k_c a_s )^{-1}$ 
takes into account that in 
Eqs.~(\ref{eq:selfnres},\ref{eq:selfares})
we have used the long-wavelength approximation (\ref{eq:Fphiphilong})
for $F^{\phi \phi}_P$.
For small $x$ and $| z |$
the asymptotic expansion of $C ( x, z )$  is
 \begin{eqnarray}
 \pi \, C ( x , z ) & = & \ln ( M^2 /4)  +
(z/2)  \ln (- 1 /z ) + O ( z, x^2),  \hspace{7mm}
 \label{eq:Fasym}
\end{eqnarray}
so that  after analytic continuation
$ z \rightarrow y + i 0$ we obtain for the imaginary part
${\rm Im} \, C ( x , y + i 0 ) \approx \Theta ( y ) y/2$ for small $x$ and $y$.
The non-analytic term  $z \ln z$ in Eq.~(\ref{eq:Fasym})
can also be derived 
directly from Eqs.~(\ref{eq:SigmaKGauss},\ref{eq:DeltaKGauss})
 by simple power-counting, which  reveals
for $D < 3$ an even stronger algebraic singularity.  
Although our explicit calculation is only valid in the
BEC regime,  the  power counting analysis of 
Eqs.~(\ref{eq:SigmaKGauss},\ref{eq:DeltaKGauss})
shows  that the coupling of the fermions to the gapless BA mode prohibits the
existence of well-defined quasi-particles for $D \leq 3$
 in the entire range of the BCS-BEC crossover.
This nicely fits to the fact that
in the interacting Bose gas
the Bogoliubov fixed point is unstable in dimensions $D \leq 3$, see
Ref.~[\onlinecite{Castellani97}].

Given Eqs.~(\ref{eq:selfnres},\ref{eq:selfares}), we may calculate
the normal component of the spectral function
 $\rho ( \bd{k} , \omega ) =  - \pi^{-1} {\rm Im} \, B ( \bd{k} , \omega + i 0 )$,
where $B ( K ) = B ( \bd{k} , i \omega )$ is the upper diagonal element
of the matrix-propagator  $ {\bf{G}}$ defined in Eq.~(\ref{eq:Gexactdef}).
From the derivation of
Eqs.~(\ref{eq:selfnres},\ref{eq:selfares}) it is clear that these
expressions are only valid  for $ | \bd{k} | \ll \sqrt{ 2 m | \mu |}$ and
 $ \bigl| | \omega | - E_{\bd{k}} \bigr| \ll | \mu |$. 
However, due to the hierarchy of energy scales $ \omega_c \ll \Delta _0 \ll | \mu |$
in the BEC regime, there are three characteristic regimes where
the spectral line-shape exhibits rather distinct behavior:

Regime I: $\Delta_0 \ll {\rm max} \{  \epsilon_{\bd{k}} ,  \bigl| 
| \omega | - E_{\bd{k}} \bigr| 
\}
 \ll | \mu  |$. In this regime the spectral function exhibits well-defined quasi-particle peaks
at $ \omega = \pm E_{\bd{k}}$, both of which have approximately the same height.
The quasi-particle damping is
$\gamma_{\bd{k}} =   2 \epsilon_s \lambda^3 ( |{\bd{k} } | a_s )^4$, which is
small compared with  $E_{\bd{k}}$, so that
the fermionic single-particle excitations
are well-defined quasi-particles.

Regime II: $\omega_c \ll {\rm max} \{ \epsilon_{\bd{k}}  , 
\bigl| | \omega | - E_{\bd{k}} \bigr|
 \}   \ll \Delta_0$.  In this intermediate regime the weight
of the negative frequency peak is a
factor of $\lambda^2$ smaller than the weight of the positive frequency peak,
which agrees with the Hartree-Fock result.
There is a considerable asymmetry between the positive and the negative
frequency part.
The quasi-particle damping in this regime is
small and momentum-independent,
 $\gamma_{\bd{k}} \approx 2 \Delta_0 \lambda^4$.

\begin{figure}[tb]
  \centering
  \epsfig{file=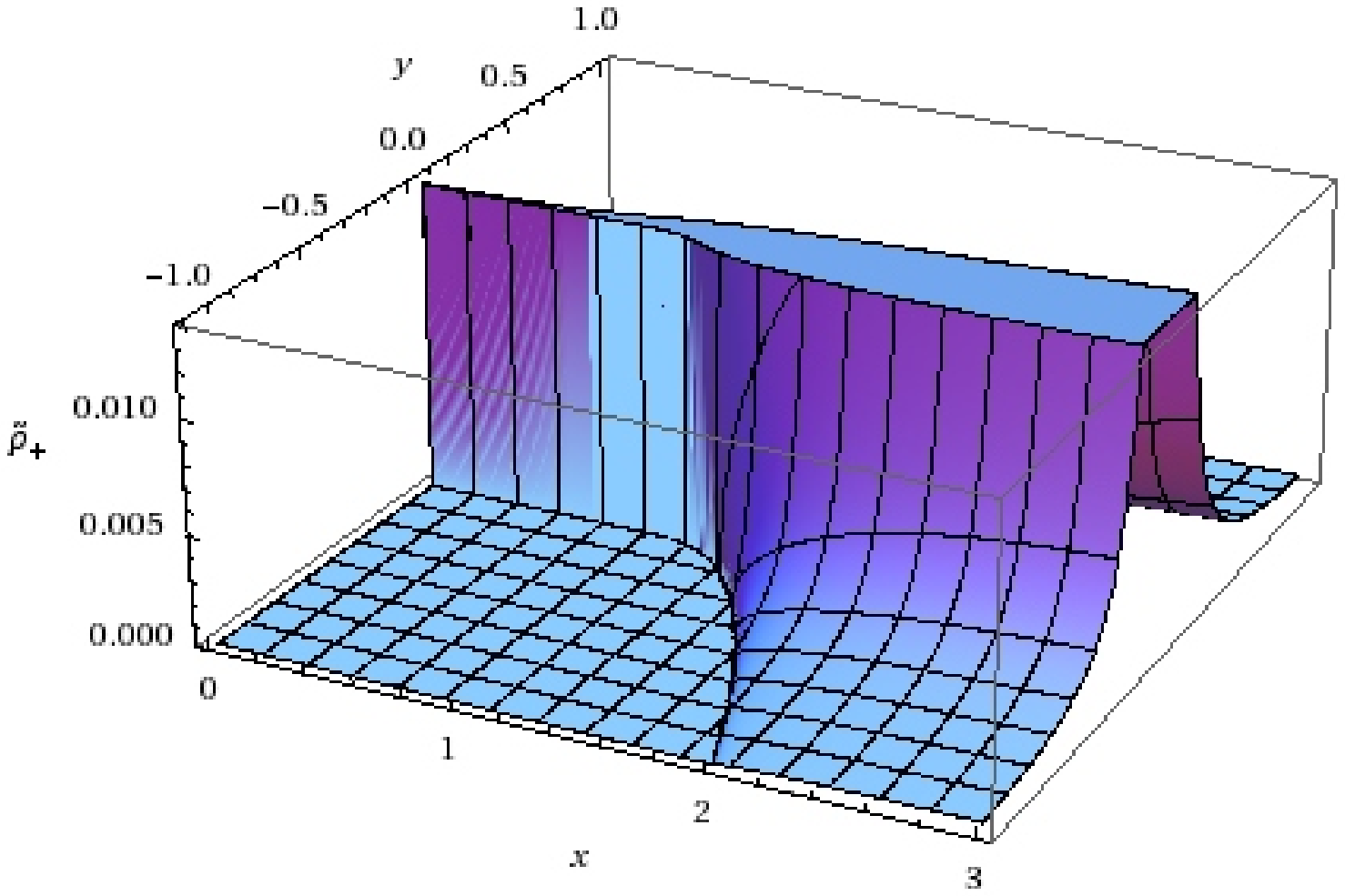,width=85mm}
 \epsfig{file=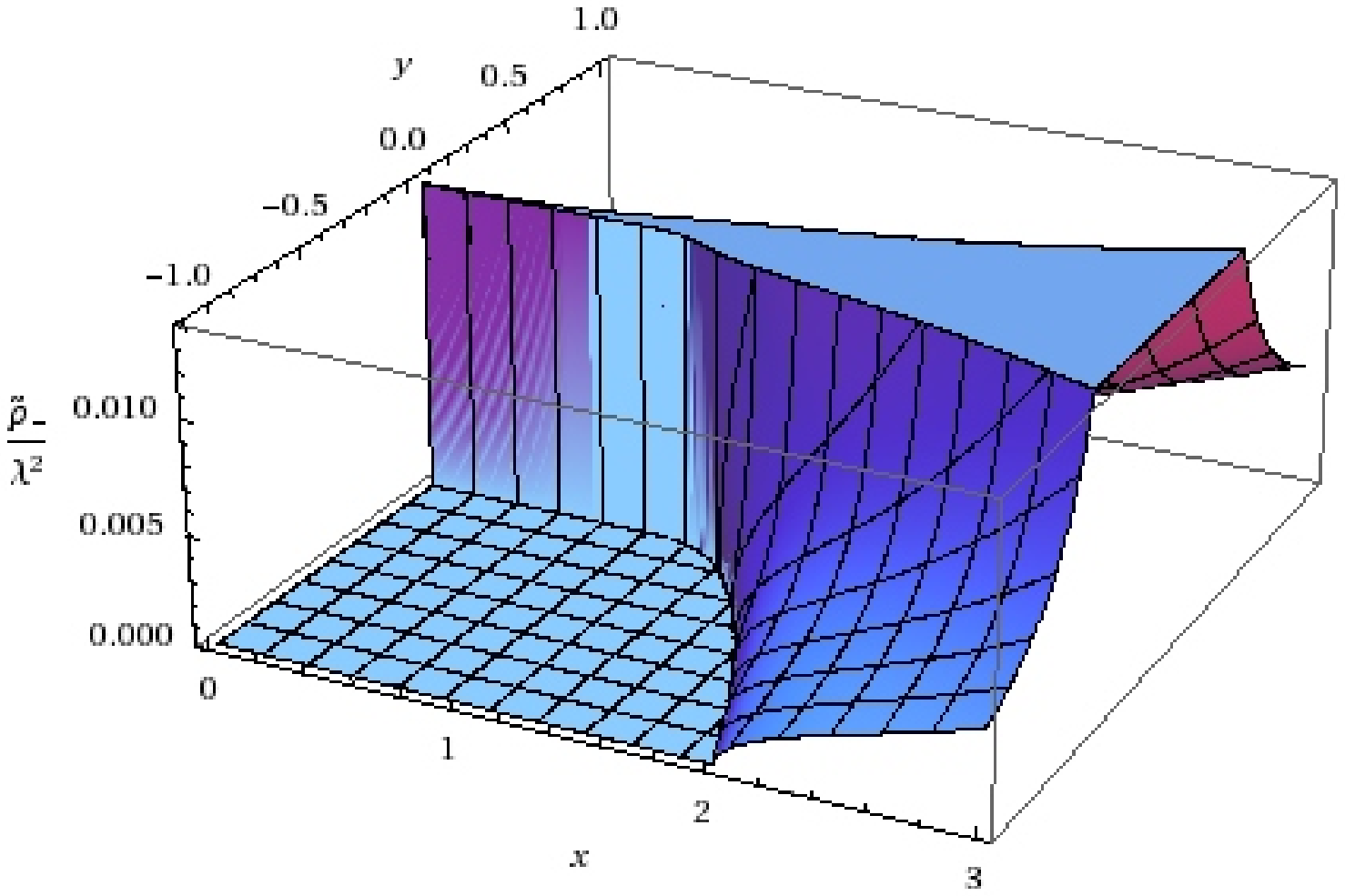,width=85mm}
  \vspace{-4mm}
  \caption{%
(Color online)
Scaling functions $\tilde{\rho}_{\pm} ( x , y )$ 
of the single-particle spectral function in the regimes II and III defined in the text.
The plots are for $\lambda = 8 \omega_c / \Delta_0 = 0.05$.
In the regime III both $ x$ and $ | y |$ are small compared with unity. 
For better comparison with $\tilde{\rho}_+ ( x , y )$   we have divided 
$\tilde{\rho}_- ( x , y )$  by $ \lambda^{2}$.
}
  \label{fig:rhoplus}
\end{figure}

Regime III: $ {\rm max} \{  \epsilon_{\bd{k}}  , \bigl| | \omega | - E_{\bd{k}} \bigr|
 \}   \ll \omega_c$.  Here it is natural to measure momenta
and energies in units of the natural scales $k_c$ and $\omega_c$
associated with the BA mode.
Writing 
$\rho ( \bd{k} , \omega )  = \Theta ( \omega ) \rho_{+} + \Theta ( - \omega ) \rho_-$,
the positive and negative frequency part of the spectral function
can be written in scaling form,
 $ \rho_{\pm} ( \bd{k} , \omega )
 = \omega_c^{-1} \tilde{\rho}_{\pm} (x  , y )$, with
 $x = |{\bd{k}}|/k_c$ and
$y =    ( | \omega | - E_{\bd{k}} ) / \omega_c$.
Graphs of the scaling functions
$\tilde{\rho}_{\pm} (x  , y )$ are shown in Fig.~\ref{fig:rhoplus}.
%
%
%
%
Note that in regime III, where $x$ and $|y|$ are both small compared with unity,
there is no spectral weight for $ | \omega | < E_{\bd{k}}$.  More generally,
for negative $y$ and $ x < 1 + \sqrt{ | y |}$
the spectral function vanishes because
$ {\rm Im} \,  C ( x , y + i 0 ) =0$ in this regime.
Physically, this is due to the kinematic constraint that a quasi-particle with momentum
$\bd{k}$ can only
spontaneously decay into another quasi-particle and a  BA quantum  if 
its  velocity $| \bd{k} | / m_{\ast}$ exceeds the velocity $c$  of the BA mode.
The scaling  functions for $ x , | y | \ll 1$ can be  written as
 $\tilde{\rho}_{+} ( x , y )  =  \tilde{\rho} ( y )$ and 
 $\tilde{\rho}_{-} ( x , y )    =   [ \lambda^2 + (y/8)^2 ] \tilde{\rho} ( y )$, where
 \begin{equation}
 \tilde{\rho} ( y )  = \Theta (y )  \pi^{-1}
Z ( y )    \gamma ( y ) /  [   
 y^2  + \gamma^2 ( y )  ],
 \label{eq:hatrhoLor}
\end{equation}
with
$Z ( y ) = 1/[ 1 -  \tilde{\lambda} \ln y ]$ and the damping function
$\gamma ( y ) = \pi \tilde{\lambda} y /  [ 1 - \tilde{\lambda} \ln y ]$.
The effective coupling parameter 
$\tilde{\lambda} = 8 \lambda^3/ \pi$ is shown in Fig.~\ref{fig:deltacc} 
as a function of $- \tilde{g}^{-1}$.
For $\omega \rightarrow E_{\bd{k}}$
the damping $\gamma ( y ) $ is only logarithmically smaller than 
the real part of the quasi-particle energy while the quasi-particle residue $Z (y )$ vanishes 
logarithmically. Such a behavior resembles  the marginal Fermi liquid scenario
postulated phenomenologically in Ref.~[\onlinecite{Varma89}].

The spectral function (\ref{eq:hatrhoLor}) implies 
a logarithmic reduction of the  density of states
$\nu (  \omega ) =
 \int \frac{d^3 k}{ (2 \pi)^3} \rho ( \bd{k} , \omega )$ for $ \omega \rightarrow 
E_0 \approx   | \mu |  + \Delta_0^2 / \epsilon_s$. For positive frequencies and  
$ \tilde{\lambda}  \ln [ \omega_c / (\omega - E_0 )  ] \gg 1$ we obtain
\begin{equation}
 \frac{\nu ( \omega ) }{\nu_0} \sim \Theta ( \omega - E_0) 
 \sqrt{ \frac{  \omega  - E_0 }{\epsilon_F } } \left[
 \tilde{\lambda} \ln \left( \frac{ \omega_c}{ \omega - E_0 } \right) \right]^{-1}.
 \label{eq:dosres} 
\end{equation}
The first factor is the
usual square-root singularity in $D=3$.
For $ \omega \rightarrow - E_0$,
the asymptotic limit of $\nu ( \omega )$
is a factor of $\lambda^2$ smaller than for positive frequencies and involves the same
logarithmic suppression.

In summary, we have shown
that in $D \leq 3$
the coupling of neutral fermions to the gapless BA mode
prohibits the existence of well-defined fermionic quasi-particles
in the superfluid state.
Although our explicit calculation is only controlled in the BEC limit,
we have argued that it remains qualitatively valid for the entire
BCS-BEC crossover, because
the non-analyticities leading to
the breakdown of the quasi-particle picture  are due to 
phase-space restrictions in the infrared regime.
In particular,
close to the unitary point $1/ \tilde{g} = 0$, 
where all energy scales are of the order of $\epsilon_F$,
we predict   a logarithmic suppression of the
density of states at the scale  $\epsilon_F$
which should be observable via tunneling experiments or other experimental
probes of the density of states in ultracold gases of neutral fermions.

This work was  supported by the DFG via SFB/TRR49.


\begin{thebibliography}{99}

\bibitem{Bloch07}
For a recent review see I. Bloch, J. Dalibard, and W. Zwerger, 
arXiv:0704.3011.
%
\bibitem{Eagles69}
D. M. Eagles, Phys. Rev.  {\bf{186}}, 456 (1969).
%
\bibitem{Leggett80}
A. J. Leggett, in {\it{Modern Trends in the Theory of Condensed Matter}},
edited by A. Pekalski and R. Przystawa, Lecture Notes
in Physics 115 (Springer, Berlin, 1980).
%
\bibitem{Nozieres85}
P. Nozi\`{e}res and S. Schmitt-Rink, J. Low Temp. Phys. {\bf{59}},
195 (1985).
%
\bibitem{Drechsler92}
M. Drechsler and W. Zwerger, Ann. Phys. (Leipzig) {\bf{1}}, 15 (1992).
%
\bibitem{Melo93}
C. A. R. S\'{a} de Melo, M. Randeria, and J. R. Engelbrecht, Phys. Rev. Lett. {\bf{71}}, 3202 (1993).
%
\bibitem{Randeria95}
M. Randeria, in {\it{Bose-Einstein Condensation}}, edited by
A. Griffin, D. Snorke, and S. Stringari, (Cambridge University Press, Cambridge, 1995).
%
\bibitem{Engelbrecht97}
J. R. Engelbrecht, M. Randeria, and C. A. R. S\'{a} de Melo,
Phys. Rev. B {\bf{55}}, 15153 (1997).
%
\bibitem{Alexandrov93}
A. S. Alexandrov and S. G. Rubin, Phys. Rev. B {\bf{47}}, 5141 (1993).
%
\bibitem{Stoof93}
H. T. C. Stoof, Phys. Rev. B {\bf{47}}, 7979 (1993).
%
%
\bibitem{Palo99}
S. De Palo, C. Castellani, C. Di Castro, and B. K. Chakraverty,
Phys. Rev. B {\bf{60}}, 564 (1999).
%
\bibitem{Hu06}
H. Hu, X.-J. Liu, and P. D. Drummond, Europhys. Lett. {\bf{74}}, 574 (2006);
H. Hu, P. Drummond, and X.-J. Liu, Nature Physics {\bf{3}}, 469 (2007).
%
\bibitem{Veillette06}
M. Y. Veillette, D. E. Sheehy, and L. Radzihovsky, Phys. Rev. A {\bf{75}}, 043614 (2007).
%
\bibitem{Nikolic07}
P. Nikoli\'{c} and S. Sachdev, Phys. Rev. A {\bf{75}}, 033608 (2007).
%
\bibitem{Diener07}
R. B. Diener, R. Sensarma, and M. Randeria, arXiv:0709.2653.
%
\bibitem{Haussmann07}
R. Haussmann, W. Rantner, S. Cerrito, and W. Zwerger,
Phys. Rev. A {\bf{75}}, 023610 (2007).
%
\bibitem{Diehl07}
S. Diehl, H. Gies, J. M. Pawlowski, and C. Wetterich,
Phys. Rev. A {\bf{76}}, 021602(R) (2007); cond-mat/0703366.
%
\bibitem{Schrieffer83}
See, for example, J. R. Schrieffer, {\it{Theory of Superconductivity}},
(Benjamin, Reading, revised printing 1983).
%
\bibitem{Varma89}
C. M. Varma, P. B. Littlewood, S. Schmitt-Rink, E. Abrahams, and 
A. E. Ruckenstein, Phys. Rev. Lett. {\bf{63}}, 1996 (1989).
%
\bibitem{footnotelabels}
We denote by  $K = ( {\bd{k}} , i \omega )$ and $P = ( {\bd{p}} , i \bar{\omega}  )$
collective labels
for momenta and  Matsubara frequencies.
Integration symbols are
  $\int_K = (\beta V )^{-1} \sum_{ {\bd{k}}, \omega } $ and
  $\int_P = (\beta V )^{-1} \sum_{ {\bd{p}}, \bar{\omega} } $
where $V $ is the volume and $\beta$ is the inverse
temperature. The corresponding
normalization of the delta-symbols is 
$\delta_{K,K'}=\beta V
\delta_{\omega,\omega'}\delta_{\bd{k},\bd{k}'}$.
%
\bibitem{Castellani97}
C. Castellani, C. Di Castro, F. Pistolesi, and G. C. Strinati,
Phys. Rev. Lett. {\bf{78}}, 1612 (1997);
F. Pistolesi, C. Castellani, C. Di Castro, and G. C. Strinati,
Phys. Rev. B {\bf{69}}, 024513 (2004).
%
\bibitem{Kreisel07}
A. Kreisel, N. Hasselmann, and P. Kopietz,
Phys. Rev. Lett. {\bf{98}}, 067203 (2007).
%
\bibitem{footnotereal}
The fields $\chi$ and $\phi$ as real as functions of
space and imaginary time such that in Fourier space
$\chi_{-P} = \chi_{P}^{\ast}$ and 
$\phi_{-P} = \phi_{P}^{\ast}$.
%
\bibitem{Sauli06}
F. Sauli and P. Kopietz, Phys. Rev. B {\bf{74}}, 193106 (2006).
%
\bibitem{Marini98}
M. Marini, F. Pistolesi, and G. C. Strinati,
Eur. Phys. J. B {\bf{1}}, 151 (1998).
%
\end{thebibliography}
\end{document}